\documentclass[paper]{ieice}
\usepackage[cmex10]{amsmath}
\usepackage{amsfonts,amssymb,amsthm}
\usepackage{array}
\usepackage{graphicx}
\usepackage{subfigure}
\graphicspath{{./figure/}}
\usepackage{fixltx2e}
\usepackage{color}
\usepackage{algorithm}
\usepackage{algorithmicx}
\usepackage{algpseudocode}

\newtheorem{theorem}{\bf Theorem}
\newtheorem{lemma}{Lemma}

\newcommand{\AmSLaTeX}{%
 $\mathcal A$\lower.4ex\hbox{$\!\mathcal M\!$}$\mathcal S$-\LaTeX}
\def\BibTeX{{\rmfamily B\kern-.05em
 \textsc{i\kern-.025em b}\kern-.08em
  T\kern-.1667em\lower.7ex\hbox{E}\kern-.125emX}}
\makeatletter
\def\tmpcite#1{\@ifundefined{b@#1}{\textbf{?}}{\csname b@#1\endcsname}}%
\makeatother

\field{D}
\vol{98}
\no{10}
\title{HISTORY: An Efficient and Robust Algorithm for Noisy 1-bit Compressed Sensing}
\authorlist{%
 \authorentry{Biao Sun}{n}{labelA}
 \authorentry{Hui Feng}{n}{labelA}
 \authorentry{Xinxin Xu}{m}{labelB}
}
\affiliate[labelA]{The author is with School of Electrical Engineering and Automation, Tianjin University, Tianjin, 300072, China. Email: sunbiao@tju.edu.cn}
\affiliate[labelB]{The author is with Microsystems Technology Center, Information Science Academy of China Electronics Technology Group Corporation, Beijing, 100086, China}

\begin{document}
\maketitle

\begin{summary}
We consider the problem of sparse signal recovery from 1-bit measurements. Due to the noise present in the acquisition and transmission process, some quantized bits may be flipped to their opposite states. These sign flips may result in severe performance degradation. In this study, a novel algorithm, termed HISTORY, is proposed. It consists of Hamming support detection and coefficients recovery. The HISTORY algorithm has high recovery accuracy and is robust to strong measurement noise. Numerical results are provided to demonstrate the effectiveness and superiority of the proposed algorithm.
\end{summary}

\section{Introduction}
Compressed sensing, as introduced in \cite{candes2006near,donoho2006compressed,candes2006robust}, addresses the problem of estimating high dimensional signals from a set of relatively few linear measurements. It was demonstrated that a sparse signal can be reconstructed exactly if the measurement matrix satisfies the restricted isometric property (RIP) \cite{candes2008restricted}. It was also shown that random matrices will satisfy the RIP with high probability if the entries are chosen according to independent and identically distributed (i.i.d.) Gaussian distribution.

In practical CS architectures, the measurements must be quantized to a finite number of bits. The extreme quantization setting where only the sign is acquired is known as 1-bit compressed sensing (1-bit CS) \cite{boufounos20081}. It has become increasingly popular due to its low computational cost and  easy implementation for hardware \cite{boufounos2010reconstruction}. In 1-bit CS, measurements of a signal $x\in\mathbb{R}^N$ are computed via
\begin{equation}
	\label{eq:1bitcs}
	{y = {\rm sign}(\mathbf{A} x)},
\end{equation}
where $x\in\mathbb{R}^N$ is the signal, $\mathbf{A}\in\mathbb{R}^{M\times N}$ is the measurement matrix, $y\in\mathbb{R}^M$ is the set of 1-bit measurements, and function $\rm sign(\cdot)$ maps the signal from $\mathbb{R}^N$ to the Boolean cube $\mathcal{B}^M:=\{-1,+1\}^M$. Since signs of real-valued measurements are used, one loses the ability to recover the magnitude of $x$ and thus assumes that the signal has a unit norm, i.e., $\|x\|_2=1$. The 1-bit CS has been studied by many people and several algorithms have been developed to recover the sparse signals \cite{boufounos20081,boufounos2009greedy,gupta2010sample,laska2011trust,zhou20121,jian2011investigation,biao2013fast}.

Despite the attractive attributes of 1-bit CS, the major disadvantage is that measurements are susceptive to noise during both acquisition and transmission \cite{gopi2013one,jacques2013robust,plan2013one}. In the noisy scenario, the output bit is randomly perturbed from the sign of the real-valued measurement, and the so-called \emph{sign flips} seriously degrade recovery performance. {The noise model is random sign-flip with probability $\rho$, i.e.,}
\begin{equation}
{y_i = b\cdot{\rm sign}(\mathbf{A}^i x),}
\end{equation}
{where $b$ equals $-1$ with probability $\rho$, $1$ with probability $1-\rho$. $y_i$ denotes the $i^\mathrm{th}$ element of $y$, and $\mathbf{A}^i$ denotes the $i^\mathrm{th}$ row of $\mathbf{A}$.} To date, researchers have developed numerous approaches for noisy 1-bit CS. Yan \emph{et al.} \cite{yan2012robust} proposed a greedy method which detects the positions of sign flips iteratively, and recovers the signals using correct measurements. However, it requires the prior knowledge of noise level, which is often intractable in practical applications. Plan \emph{et al.} \cite{plan2013robust} proposed a constrained optimization method with a linear objective. This convex formulation can work with a general notion of noise and achieve error for both exactly and approximately sparse signals. Ai \emph{et al.} \cite{ai2014one} extends \cite{plan2013robust} to sub-Gaussian measurements, and gets an irreducible component in the error and cannot be reduced by increasing the sample size or otherwise. However, they are computationally inefficient and difficult for hardware implementation. Recently, Zhang \emph{et al}. \cite{zhang2014efficient} developed an efficient passive algorithm with closed-form solution, which improves the recovery performance for exactly $K$-sparse signals. Due to its high performance, robustness, and computational efficiency, they can be seen as the state-of-the-art algorithm for noisy 1-bit CS.

{This study focuses on recovering EXACTLY $K$-sparse signals that have $K$ nonzero coefficients in the noisy setting for 1-bit CS. We define $\Sigma_K$ to be the set of all exactly $K$-sparse signals with unit norm as}
\begin{equation}
{\Sigma_K \overset{\mathrm{def}}{=} \{v\in \mathbb{R}^N : \|v\|_0 = K,\ \|v\|_2 = 1\}.}
\end{equation}
A novel algorithm is proposed in this paper. Termed HISTORY, it consists of two key parts, namely \emph{HammIng Support deTection}, and \emph{cOefficients RecoverY}. The former aims to construct a candidate supports set by detecting possible supports of nonzero entries. The latter aims to calculate the coefficients belonging to the candidate supports set. Experimental results show that the proposed algorithm has high recovery performance than the state-of-the-art. Also, because containing no iterative step, it is computationally efficient and easy to implement.

\section{HISTORY Algorithm}
\label{sec:HISTORY_Algorithm}
The main objective of this section is to characterize the HISTORY algorithm. Notations used throughout this paper are first described, then the two key parts of HISTORY are introduced in sequence.

\subsection{Notations}
Boldfaced capital letters such as $\mathbf{A}$ are used for matrices. Italic capital letters such as $S$ denote sets. For a matrix $\mathbf{A}$, {the notations $\mathbf{A}^i$}, $\mathbf{A}_j$, {$\mathbf{A}^i_j$}, $\mathbf{A}^\mathrm{T}$, and $\mathbf{A}_S$ denote its {$i^\mathrm{th}$ row}, $j^\mathrm{th}$ column, $ij^\mathrm{th}$ element, transpose, and sub-matrix that contains the columns with indices in $S$, respectively. Small letters such as $x$ are reserved for vectors and scalars. {A vector $x$ is called exactly $K$-sparse if $K$ of its coefficients are nonzero.} For a vector $x$, $x_j$, $\|x\|_p$, and $x_S$ denote the $j^\mathrm{th}$ element of the vector, its $p$-norm, and sub-vector that contains the elements with indices in $S$, respectively. For two vectors $u\in\mathbb{R}^N$ and $v\in\mathbb{R}^N$, the notation $H(u,v)$ denotes the Hamming distance between them, which is defined as
\begin{equation}
	H(u,v)\overset{\mathrm{def}}{=}\#\left(u_j\neq v_j\right),\quad j\in1,2,\dots,N.
\end{equation}
For an event $E$, the notation $\mathbb{P}(E)$ denotes its probability. {For a random variable $a$, the notations $\mathbb{E}(a)$ and $\mathbb{D}(a)$ denote its expectation and variance, respectively.}

\subsection{Hamming support detection}
To detect possible supports of nonzero coefficients from noisy 1-bit measurements, a \emph{Hamming support detection} method is developed based on Angle Proportional Probability (APP), which is outlined as follows.
\begin{theorem}[Angle Proportional Probability]
	\label{Theorem:APP}
	Let $x\in\Sigma_K$ be an {exactly} $K$-sparse signal with $\|x\|_2=1$. Let $\phi$ be a Gaussian random vector which is drawn uniformly from the unit $\ell_2$ sphere in $\mathbb{R}^N$ (i.e., each element of $\phi$ is firstly drawn i.i.d. from the standard Gaussian distribution $\mathcal{N}(0,1)$. Define an event $E$ to be
	\begin{equation}
		\label{eq:event}
		E: {\rm sign}(x^{\rm T}\phi)\neq{\rm sign}(\phi_j),
	\end{equation}
	then it holds,
	\begin{equation}
		\label{eq:APP}
		\mathbb{P}(E) = \frac{1}{\pi}{\rm arccos}(x_j).
	\end{equation}
\end{theorem}
The proof can be found in Appendix A. In particular, it shows that $\mathbb{P}(E)$ has a cosine function relationship with the $j$-th element of $\phi$. Thus, $x_j$ can be uniquely identified by $\mathbb{P}(E)$. In addition, the probability can be estimated from the instances of the random variable $\mathrm{sign}(x^{\rm T}\phi)$, which are exactly the 1-bit measurement vector $y$ defined in (\ref{eq:1bitcs}). Therefore, $y$ contains sufficient information to reconstruct $x_j$ from the estimation of $\mathbb{P}(E)$.

In the noisy setting, due to the fact that the signs of $y$ are randomly perturbed, $x_j$ cannot be computed directly from (\ref{eq:event}) and (\ref{eq:APP}). However, given the noise level (sign flip ratio) as a prior knowledge, we have the following lemma.
\begin{lemma}
	\label{Lemma:sign-flip}
	Given an {exactly} $K$-sparse signal {$x\in\Sigma_K$} with $\|x\|_2=1$, a standard Gaussian measurement matrix $\mathbf{A}\in\mathbb{R}^{M\times N}$, and a 1-bit measurements vector $y = {\rm sign}(\mathbf{A} x)$. In the noisy setting, suppose the sign flip ratio $\rho<0.5$, define $P\in[0,1]^N$ as a probability vector with $P_j$ denoting its $j$-th element as
	\begin{equation}
		P_j\overset{\mathrm{def}}{=}\mathbb{P}\left(\mathrm{sign}\left(y_i\right)\neq\mathrm{sign}\left({\mathbf{A}^i_j}\right)\right),
	\end{equation}
	and it holds
	\begin{equation}
		\label{eq:hamming_rho}
		P_j = \frac{1-2\rho}{\pi}{\rm arccos}(x_j)+\rho.
	\end{equation}
\end{lemma}
The proof can be found in Appendix B. 
{From Lemma \ref{Lemma:sign-flip}, we note that the Hamming distance between $y$ and $\mathbf{A}_j$ obeys the binomial distribution, i.e.,
\begin{equation}
H\{y,\mathbf{A}_j\} \sim B(M,P_j).
\end{equation}
Moreover, by the definition of binomial distribution, we have
\begin{equation}
\mathbb{E}\big(H\{y,\mathbf{A}_j\}\big) = M\left(\frac{1-2\rho}{\pi}{\rm arccos}(x_j)+\rho\right).
\end{equation}
}
Consequently, given the noise level and a relatively high measurement dimension, $P_j$ can be well estimated by computing the Hamming distance, then $x_j$ can be estimated accordingly. However, directly estimating $x_j$ from (\ref{eq:hamming_rho}) is intractable. For one thing, with the decrease of measurement dimension, the coefficients estimation performance degrades significantly. For another, (\ref{eq:hamming_rho}) requires the sign flip ratio $\rho$ as prior knowledge, which is often unknown in practical applications. To address the first problem, we only detect possible supports in current part and leave the coefficients estimation to the next one. {To address the second problem, we propose the following lemma.
\begin{lemma}
	\label{Lemma:error_bound}
	Given a finite measurement dimension $M$ and a sign flip ratio $\rho<0.5$, for any two different elements of $x$, denoted by $x_u, x_v, u\neq v$, if $x_u-x_v>\epsilon$, where $\epsilon$ is a small positive constant, we have
	\begin{equation}
	\mathbb{P}\big(H\{y,\mathbf{A}_u\}<H\{y,\mathbf{A}_v\}\big)\geq 1+C_1-C_2\epsilon^{-2},
	\end{equation}
	where $C_1$ and $C_2$ are constants and
	\begin{equation}
	\begin{split}
	C_1 &= \frac{1}{4M}, \\
	C_2 &= \frac{\pi^2}{4M(1-2\rho)^2}. \\
	\end{split}
	\end{equation}
\end{lemma}
The proof can be found in Appendix C. Note that with the increase of $M$, the probability $\mathbb{P}\big(H\{y,\mathbf{A}_u\}<H\{y,\mathbf{A}_v\}\big)$ also increases, and when $M\rightarrow\infty$, we have
\begin{equation}
\mathbb{P}\big(H\{y,\mathbf{A}_u\}<H\{y,\mathbf{A}_v\}\big)\rightarrow 1, \quad \forall\ x_u-x_v>\epsilon.
\end{equation}}

{From Lemma \ref{Lemma:error_bound}, it is easy to verify that despite the value of $\rho$, $P_j$ in (\ref{eq:hamming_rho}) is a monotone decreasing function with respect to $x_j$.} The main point is that despite the noise level, the amplitude order of nonzero coefficients will maintain, while the dependencies in $\rho$ will vanish in the corresponding Hamming distance. Therefore, we can set $\rho$ to be an arbitrary value (e.g. $\rho = 0$) and compute approximate amplitudes of each coefficient via (\ref{eq:hamming_rho}), then form the candidate supports set by selecting the supports with largest amplitudes.

\subsection{Coefficients recovery}
Providing the candidate supports set, denoted by $S$, the next part is coefficients recovery, which aims to compute the amplitudes of nonzero coefficients. In this paper, we try to compute the coefficients vector $c$ by solving the following constrained least squares problem,
\begin{equation}
	\label{eq:ls}
	c^* = \underset{c\in\mathbb{R}^{|S|}}{\mathrm{minimize}}\|y-\mathbf{A}_S\cdot c\|_2\quad \mathrm{s.t.} \quad \|c\|_0 = K,
\end{equation}
where $\|c\|_0$ denotes the $0$-norm of $c$, i.e., counting the number of nonzero coefficients in $c$. Note that (\ref{eq:ls}) is an overdetermined system when ${|S|<M}$. Thus, the sparsest solution to (\ref{eq:ls}) is given by
\begin{equation}
	\label{eq:mldivide}
	c^* = \mathbf{A}_\mathcal{S} \setminus y,
\end{equation}
where ``$\setminus$" denotes the \emph{left matrix divide} operation. (\ref{eq:mldivide}) can be solved via the QR decomposition \cite{trefethen1997numerical} efficiently.

Based on the two parts described above, the HISTORY algorithm is fully summarized in Algorithm \ref{alg:HISTORY}, {where $\mathrm{abs}(h)$ denotes the absolute value of each element of the vector $h$}, FindSupp$\big({\mathrm{abs}(h)},\alpha K\big)$ returns the supports of the largest $\alpha K$ elements in ${\mathrm{abs}(h)}$, and $\mathrm{H_K}(\cdot)$ denotes the hard-thresholding operator who only preserves the largest $K$ coefficients {in magnitude} and set others to 0. {$\alpha$ is a parameter that controls the redundancy of support detection. For $\alpha>1$, Algorithm \ref{alg:HISTORY} first selects more than $K$ supports to form the candidate set. After computing the coefficients vector, the final $K$-sparse solution is obtained by hard-thresholding as in step 10. Note that when $M$ is small, a high $\alpha$ is necessary to ensure the support detection accuracy. With the increase of $M$, a small $\alpha$ is sufficient to detect the supports accurately. In addition, a small $\alpha$ can decrease the computational complexity of (\ref{eq:ls}), thus boost the whole algorithm. Based on the above analysis, we propose selecting $\alpha$ adaptively as}
\begin{equation}
\label{eq:alpha_adaptive}
{\alpha = 1+\alpha_0e^{-\tau\frac{M}{N}},}
\end{equation}
{where $\alpha_0$ is the initial quantity and $\tau$ is the exponential decay constant.}

\begin{algorithm}[ht]
	\caption{HISTORY}
	\label{alg:HISTORY}
	\begin{algorithmic}[1]
		\Require
		$y,\mathbf{A},K,\alpha$
		\State \textbf{Initialize:} $x^* = \mathrm{Zeros}(N)$
		\For{each $j\in 1,\dots,N$}
		\State $P_j = H\{y,\mathbf{A}_j\}/M$
		\State $h_j = \text{cos}(\pi P_j)$
		\EndFor
		\State $S = \mathrm{FindSupp}\big({\mathrm{abs}(h)},\alpha K\big)$
		\State $c^* = \mathbf{A}_\mathcal{S} \setminus y$
		\State $x^*_S = c^*$
		\If{$\alpha>1$}
		\State ${x^* = \text{H}_\text{K}(x^*)}$
		\EndIf
		\State $x^* = x^*/\|x^*\|_2$
		\Ensure
		recovered sparse signal $x^*$
	\end{algorithmic}
\end{algorithm}

It is worth noting that Algorithm \ref{alg:HISTORY} is a nearly-linear time algorithm, with its computational complexity to be $O(MN)$. Therefore, the proposed algorithm runs significantly faster than iterative algorithms.

\section{Experiments}
\label{sec:Results}
\subsection{Experimental Setup}
The target vector $x\in\mathbb{R}^N$ is generated by drawing its nonzero elements from the standard Gaussian distribution, and then normalized to have unit norm. The locations of the $K$ nonzero coefficients of $x$ are randomly selected. The elements in the measurement matrix $\mathbf{A}\in\mathbb{R}^{M\times N}$ are also drawn from the standard Gaussian distribution. To generate sign flips, the measurement vector $y$ is firstly acquired as in (\ref{eq:1bitcs}), then the sign of every element in $y$ is flipped with probability $\rho$. For each setting of $M$, $N$, $K$, and $\rho$, the recovery experiment is repeated for 100 trials, and the average recovery error, denoted by $\|x-x^*\|_2/\|x\|_2$, is reported. {In all experiments, The parameter $\alpha$ is selected adaptively as in (\ref{eq:alpha_adaptive}) with $\alpha_0=4$ and $\tau=1$.}

The HISTORY algorithm is compared with the following three algorithms,
\begin{itemize}
	\item BIHT-$\ell_2$: a heuristic algorithm proposed in \cite{jacques2013robust}, which has been proved to have better performance than BIHT in the noisy setting. The maximum iterative number and step size are set to 200 and 1, respectively \footnote{A matlab implementation of BIHT-$\ell_2$ algorithm can be downloaded from http://perso.uclouvain.be/laurent.jacques\\
		/index.php/Main/BIHTDemo.}.
	\item Convex: a provable algorithm proposed in \cite{plan2013robust}, which solves a convex optimization problem to recover the sparse signal \footnote{The CVX package is used to solve this optimization problem. The package can be downloaded from http://cvxr.com/cvx/.}.
	\item Passive: an efficient optimization algorithm with closed-form solution proposed in \cite{zhang2014efficient}, experimental results illustrated that their passive algorithm outperforms other baselines. The regularization parameter $\gamma$ is set to $\sqrt{\frac{\log N}{M}}$, which is the optimal choice in \cite{zhang2014efficient}.
\end{itemize}

\subsection{Results}
\begin{figure}[tb]
	\centering
	\subfigure[]{\label{fig:Exp_alpha_acc}\includegraphics[width =.5\textwidth]{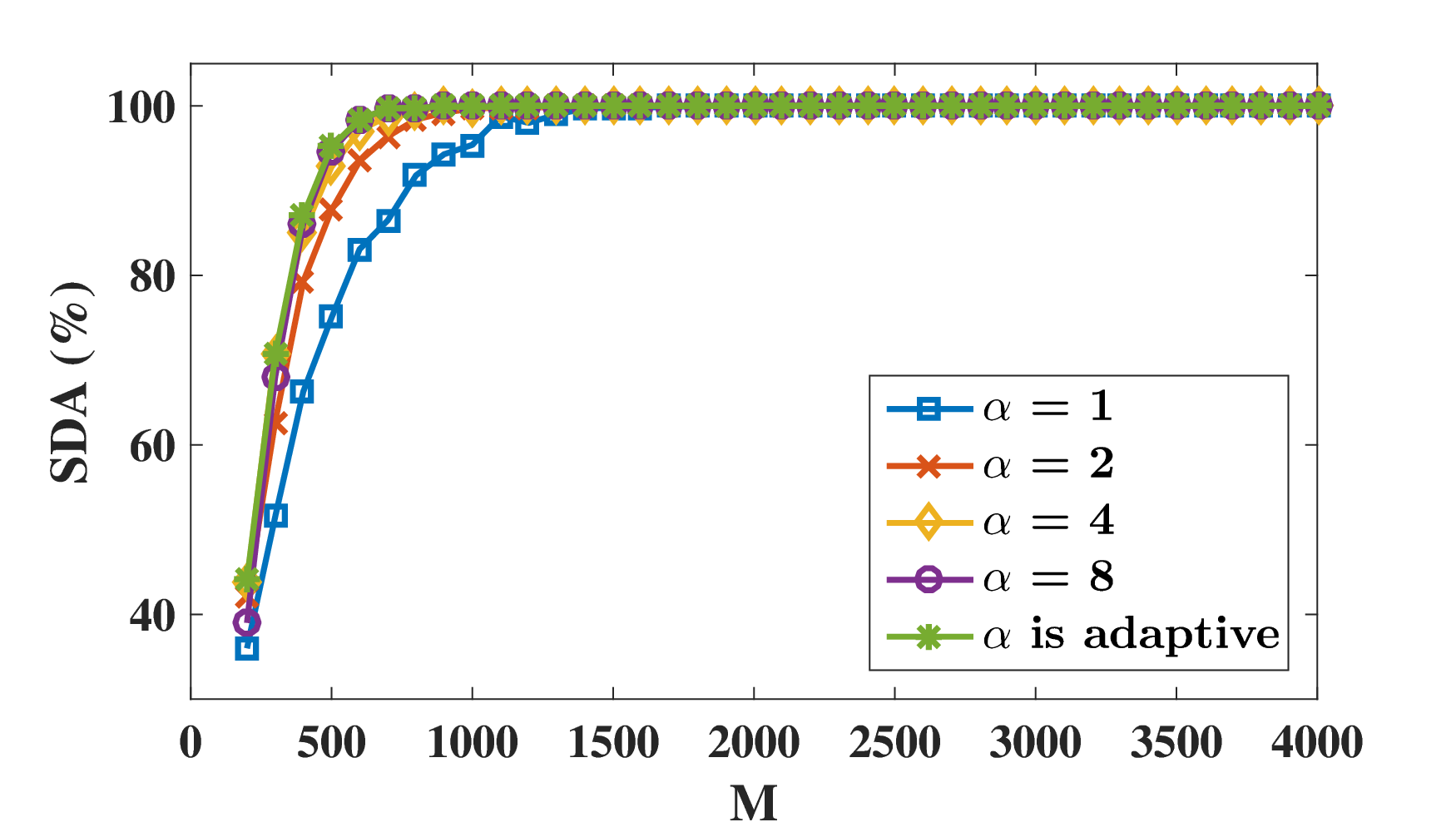}}
	\subfigure[]{\label{fig:Exp_alpha_time}\includegraphics[width =.5\textwidth]{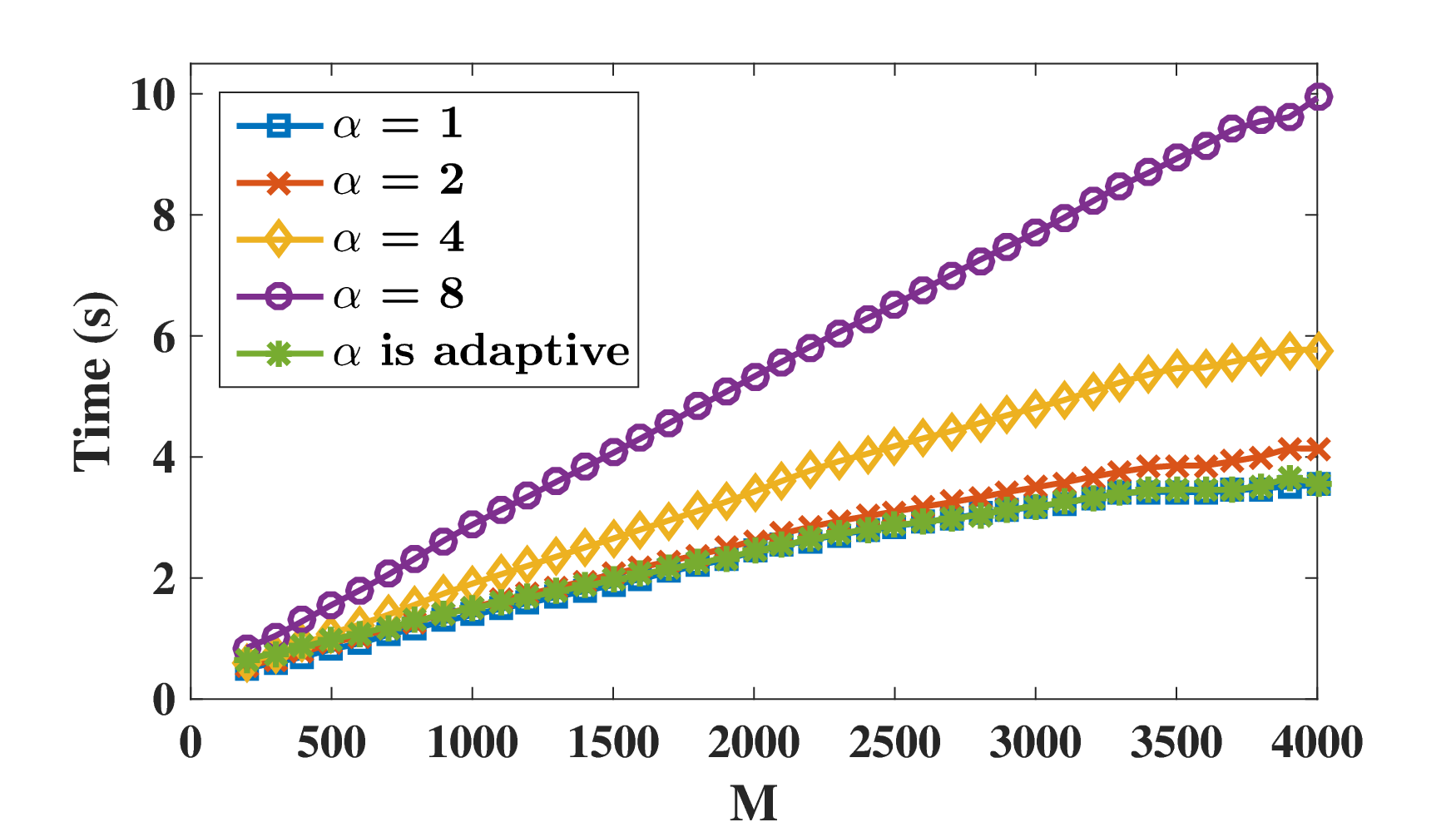}}
	\caption{(a) Evaluate support detection accuracy with different $\alpha$, (b) evaluate CPU time with different $\alpha$, when $N = 1000$, $K = 10$, $\rho=0.1$, and $M$ is varied from 200 to 4000.}
	\label{fig:Exp_alpha}
\end{figure}

{\subsubsection{Support detection accuracy and computational efficiency of the adaptive $\alpha$} Firstly, the support detection accuracy with different $\alpha$ is studied. Parameters are set as $N=1000$, $K=10$, $\rho=0.1$, and $M$ is varied from 200 to 4000. The parameter $\alpha$ is selected adaptively as in (\ref{eq:alpha_adaptive}) with parameters $\alpha_0=4$ and $\tau=1$. $\alpha$ is also selected with fixed values as $\alpha=1,2,4,8$ for comparison. The support detection accuracy (SDA) is employed to quantify the
percentage of detection success between the original supports and the reconstructed supports. The SDA is defined as}
\begin{equation}
{\mathrm{SDA} = \frac{\#\big(\mathrm{supp}(x^*)\cap\mathrm{supp}(x)\big)}{\#\big(\mathrm{supp}(x)\big)}\times 100\%.}
\end{equation}
{The support detection accuracy curve with different $\alpha$ is shown in Fig. \ref{fig:Exp_alpha_acc}. It's observed that the adaptive $\alpha$ has the highest SDA. Although $\alpha=8$ has better performance than $\alpha=1,2,4$ with small $M$, all $\alpha$ values have same SDA after $M>1500$. Therefore, a high $\alpha$ is not necessary when $M$ is large.

To study the computational efficiency of the adaptive $\alpha$, the CPU time of HISTORY with different $\alpha$ is evaluated. For each point, the recovery experiment is repeated for 100 trials, and the total cpu time is reported in Fig. \ref{fig:Exp_alpha_time}. It is observed that larger $\alpha$ costs more computational resource. HISTORY with $\alpha=8$ costs more than double CPU time of that with $\alpha=1$ to recover the signals. The adaptive $\alpha$ costs least CPU time, i.e., almost same with $\alpha=1$. Therefore, the adaptive $\alpha$ has better computational efficiency than fixed ones.}

\subsubsection{Recovery error versus measurement dimension} 
\begin{figure}[tb]
	\begin{center}
		\includegraphics[width=.5\textwidth]{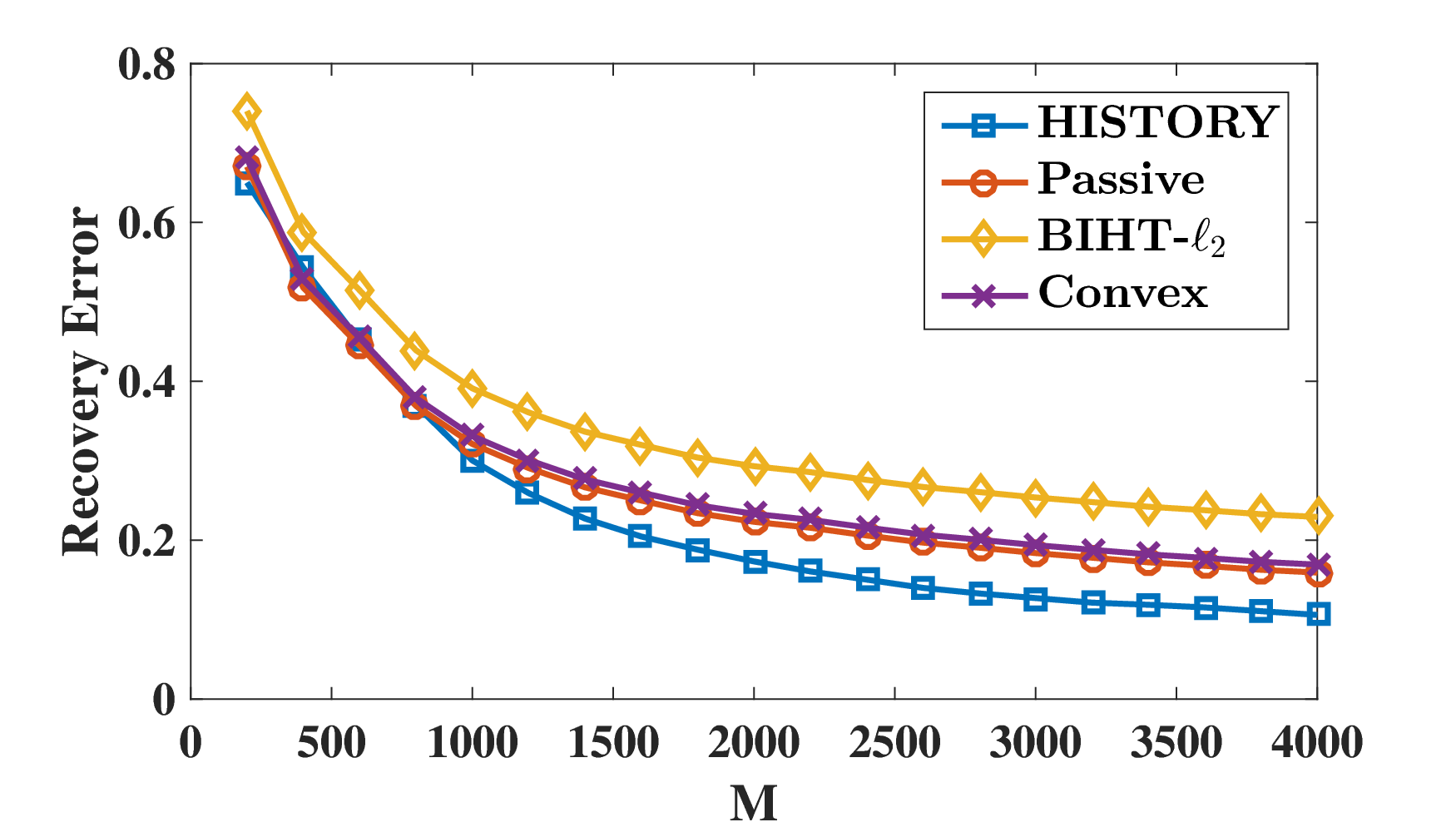}
	\end{center}
	\caption{Evaluate recovery error of each algorithm versus measurement dimension $M$, when $N = 1000$, $K = 10$, and $\rho=0.1$.}
	\label{fig:Exp_M_noisy}
\end{figure}

Then the recovery error at different measurement dimension $M$ is studied. Parameters are set as $N=1000$, $K=10$, $\rho=0.1$, and $M$ is varied from 200 to 4000. The recovery error curve is shown in Fig. \ref{fig:Exp_M_noisy}. It is observed that with the increase of $M$, the recovery errors of all algorithms decrease. In particular, BIHT-$\ell_2$ has the worst performance among these algorithms, that is because it is very sensitive to noise in the 1-bit measurements. In contrast, HISTORY has the best performance, especially when $M$ is relatively large. The recovery errors of Convex and Passive are very similar.

\subsubsection{Recovery error versus sparsity} 
\begin{figure}[tb]
	\begin{center}
		\includegraphics[width=.5\textwidth]{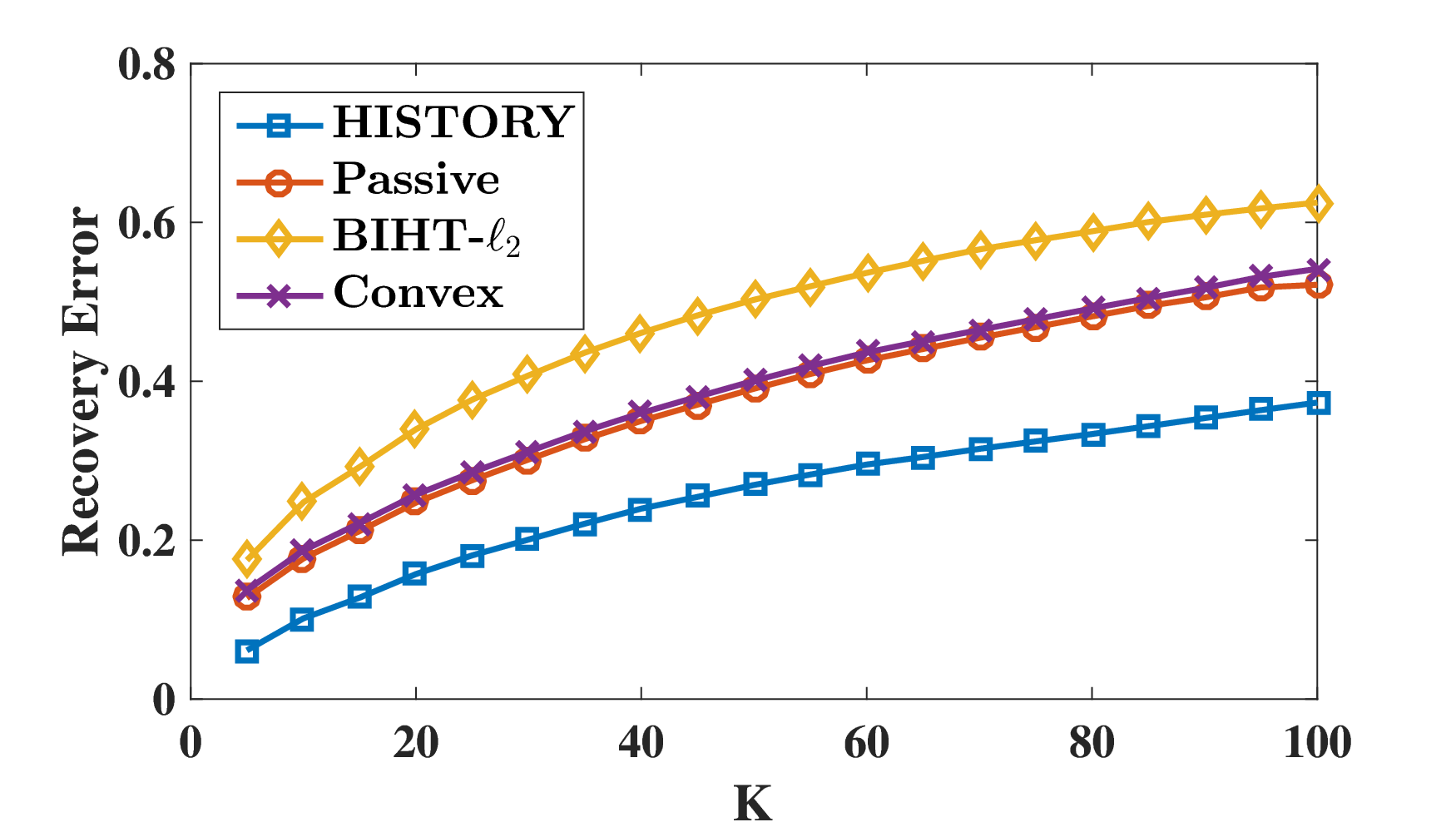}
	\end{center}
	\caption{Evaluate recovery error of each algorithm versus sparsity $K$, when $N = 1000$, $\rho = 0.1$, and $M = 4000$.}
	\label{fig:Exp_K}
\end{figure}

Then the recovery error at different sparsity $K$ is evaluated. Parameters are set as $N=1000$, $\rho=0.1$, $M=4000$, and $K$ is varied from 10 to 200. The recovery error curves are shown in Fig. \ref{fig:Exp_K}. Results show that with the increase of $K$, the recovery errors of all algorithms increase. In particular, among these algorithms, HISTORY has the best performance while BIHT-$\ell_2$ has the worst one. In addition, Passive and Convex almost have the same performance. Finally, we would like to emphasize that HISTORY increases its advantage with the increase of $K$, i.e., it is less sensitive to sparsity than other algorithms.

\subsubsection{Recovery error versus sign flip ratio} 
\begin{figure}[tb]
	\begin{center}
		\includegraphics[width=.5\textwidth]{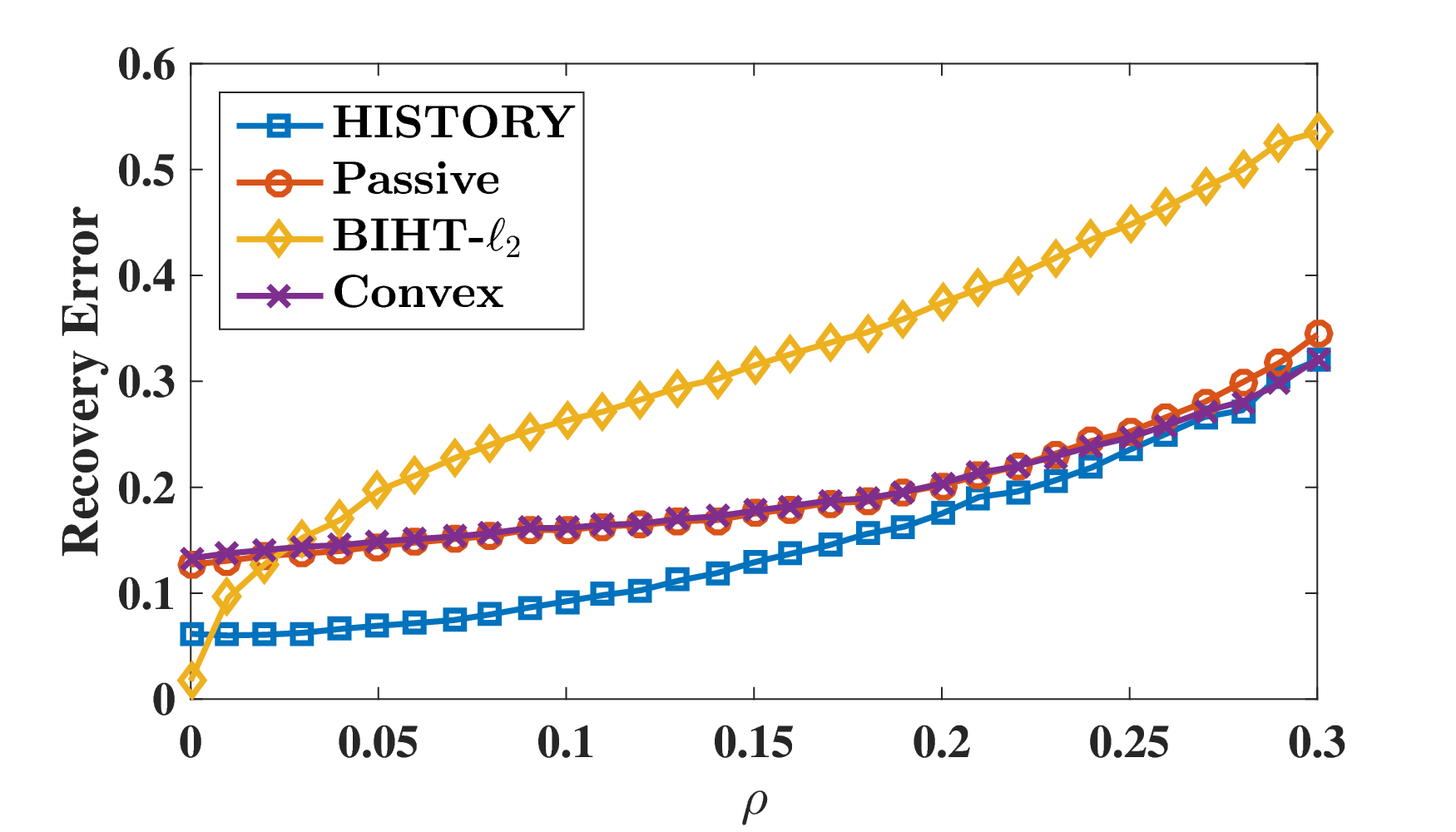}
	\end{center}
	\caption{Evaluate recovery error of each algorithm versus sign flip ratio $\rho$, when $N = 1000$, $K = 10$, and $M = 4000$.}
	\label{fig:Exp_rho}
\end{figure}

Next, the recovery error at different sign flip ratio $\rho$ is evaluated. Parameters are set as $N=1000$, $K=10$, $M=4000$, and $\rho$ is varied from 0 to 0.5. The recovery error curves are shown in Fig. \ref{fig:Exp_rho}. Though BIHT-$\ell_2$ had the minimum recovery error when $\rho$ is small, with the increase of $\rho$, its recovery error increased very quickly, making it be the worst algorithm at the high sign flip ratio. Passive and Convex had almost the same performance, which are better than that of BIHT-$\ell_2$. HISTORY has the best performance both at high and low sign flip ratio. Thus, HISTORY has the best noise robustness among these algorithms.

\subsubsection{Recovery error under misspecified model} 
\begin{figure}[tb]
	\begin{center}
		\includegraphics[width=.5\textwidth]{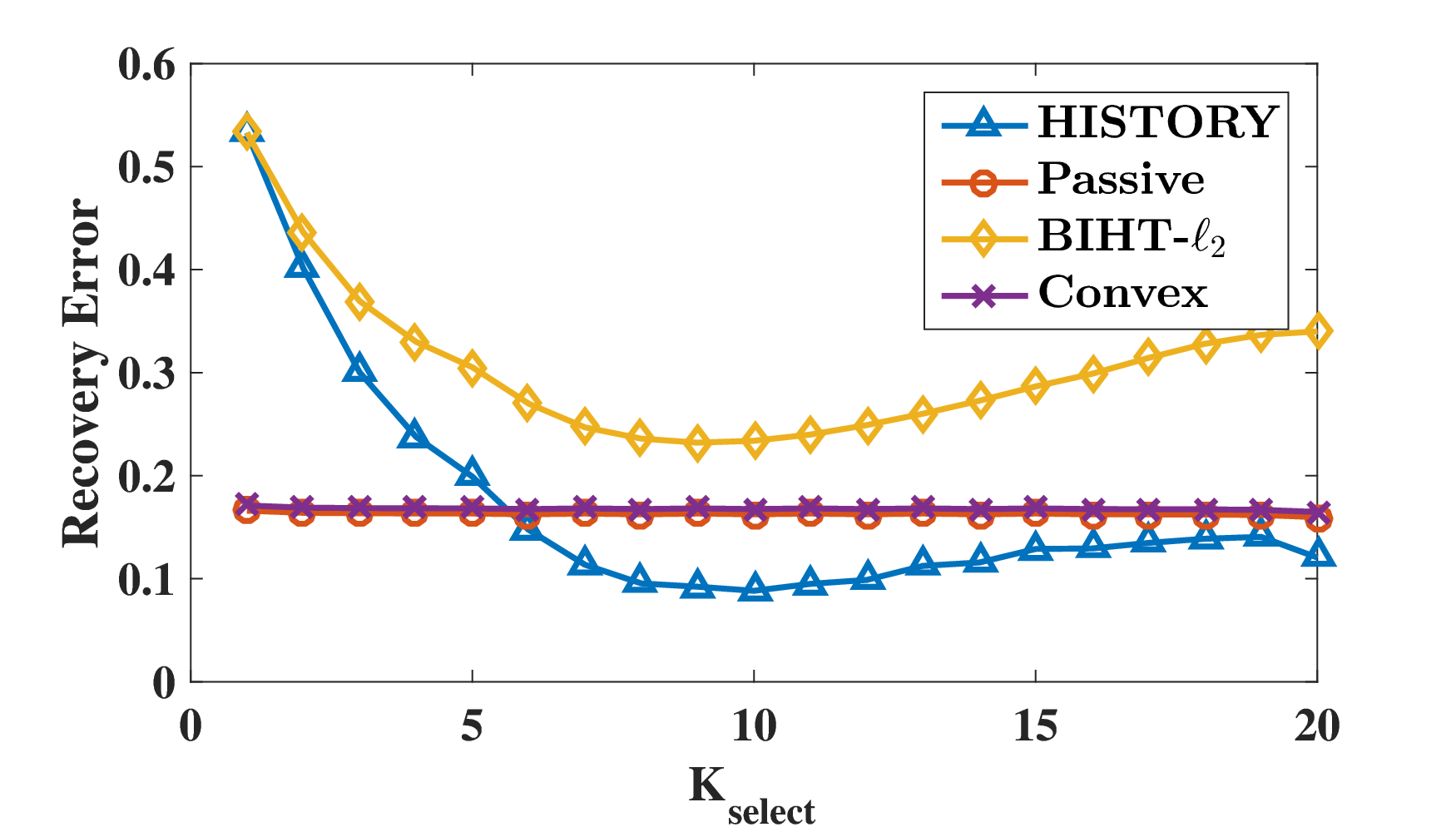}
	\end{center}
	\caption{Evaluate recovery error of each algorithm when $K$ is unknown. Parameters are set to $N = 1000$, $K = 10$, $M = 4000$, and $\rho = 0.1$. $K$ is selected from 1 to 20.}
	\label{fig:Exp_K_unknown}
\end{figure}

Next, we study the error of each algorithm under the misspecified model, i.e., the sparsity of original signal is unknown. Parameters are set as $N=1000$, $K=10$, $M=4000$, $\rho=0.1$, and we select $K_\mathrm{select}$ from 1 to 20 to evaluate the algorithms. The recovery error curves are shown in Fig. \ref{fig:Exp_K_unknown}. Results show that the recovery error of HISTORY sharply drops at the correct $K_\mathrm{select}=K$. Moreover, HISTORY performs better than Passive and Convex in a neighborhood of $K$. Under misspecification with $K_\mathrm{select}<K$, the recovery error is large since the error from unrecovered coefficients is large. For $K_\mathrm{select}>K$, the nonzero coefficients are correctly recovered so that the corresponding error is small, but there is some additional error due to noise. {To further improve the performance of HISTORY when $K$ is unknown, many approaches can be used to estimate the sparsity level. For example, we can select a regularization parameter $\gamma$ first, and then use the thresholding method to estimate $K$ as proposed in \cite{zhang2014efficient}. Other approaches such as 1-bit one scan \cite{li2015one} and sudocodes-based method \cite{ma2014two} can also be used for sparsity estimation. We will leave this part for future research.}

\subsubsection{Computational complexity} To evaluate the computational complexity of each algorithm, we study the running time of them. Parameters are set as $N=1000$, $K=10$, $M=4000$, and $\rho=0.1$. The running time of those algorithms can be found in Table \ref{table_runtime}. Results show that the running time of HISTORY and Passive are similar, while that of Convex and BIHT-$\ell_2$ are significantly higher.

\begin{table}[ht]
	\renewcommand{\arraystretch}{1.3}
	\caption{Running time of each algorithm, when $N=1000$, $K=10$, $M=4000$, and $\rho=0.1$. For BIHT-$\ell_2$, there is no formal stoping criterion, and we report the running time after 100 iterations.}
	\label{table_runtime}
	\centering
	\begin{tabular}{l|c|c|c|c}
		\hline
		\bfseries Algorithm & BIHT-$\ell_2$ & Convex & Passive & HISTORY \\
		\hline
		\bfseries Time (s) & {331.29} & {155.47} & {3.04} & {3.22} \\
		\hline
	\end{tabular}
\end{table}

\section{Conclusion}
\label{sec:Conclusion}
In this paper, we have developed an efficient and robust algorithm for noisy 1-bit compressive sensing. Compared with the existing methods, the proposed algorithm has several important advantages. It is robust to noise, it is computationally efficient, it has lower sample complexity, and it is easy to implement. Experimental results provide sound support to our theoretical development.

\section*{Acknowledgments}
This work was supported by the National Natural Science Foundation of China under Grants 61271321, 61473207 and 61401303, the Ph.D. Programs Foundation of the Ministry of Education of China under Grant 20120032110068, and Tianjin Key Technology Research and Development Program under Grant 14ZCZDS F00025.



\begin{thebibliography}{99}
	
\bibitem{candes2006near}
E. J. Candes and T. Tao, ``Near-optimal signal recovery from random projections: Universal encoding strategies?," IEEE T. Inform Theory, vol.52, no.12, pp. 5406-5425, 2006.

\bibitem{donoho2006compressed}
D. L. Donoho, ``Compressed sensing," IEEE T. Inform Theory, vol. 52, no. 4, pp. 1289-1306, 2006.
	
\bibitem{candes2006robust}
E. J. Cand{\`e}s, J. Romberg, and T. Tao, ``Robust uncertainty principles: Exact signal reconstruction from highly incomplete frequency information," IEEE T. Inform Theory, vol. 52, no. 2, pp. 489-509, 2006.
	
\bibitem{candes2008restricted}
E. J. Cand{\`e}s, ``The restricted isometry property and its implications for compressed sensing," CR Math, vol. 346, no. 9, pp. 589-592, 2008.
	
\bibitem{boufounos20081}
P. T. Boufounos and R. G. Baraniuk, ``1-bit compressive sensing," Proc. 42nd Annual Conf. on Information Sciences and Systems (CISS), Princeton, USA, pp. 16-21, 2008.
	
\bibitem{boufounos2010reconstruction}
P. T. Boufounos, ``Reconstruction of sparse signals from distorted randomized measurements," Proc. IEEE International Conf. on Acoustics Speech and Signal Processing (ICASSP), Dallas, USA, pp. 3998-4001, 2010.
	
\bibitem{boufounos2009greedy}
P. T. Boufounos, ``Greedy sparse signal reconstruction from sign measurements," Proc. 43rd Asilomar Conference on Signals, Systems and Computers, Asilomar, USA, pp. 1305-1309, 2009.
	
\bibitem{gupta2010sample}
A. Gupta, R. Nowak, and B. Recht, ``Sample complexity for 1-bit compressed sensing and sparse classification," Proc. IEEE International Symposium on Information Theory (ISIT), Austin, USA, pp. 1553-1557, 2010.
	
\bibitem{laska2011trust}
J. N. Laska, Z. Wen, W. Yin, and R. G. Baraniuk, ``Trust, but verify: Fast and accurate signal recovery from 1-bit compressive measurements," IEEE T. Signal Proces, vol. 59, no. 11, pp. 5289-5301, 2011.
	
\bibitem{zhou20121}
T. Zhou and D. Tao, ``1-bit hamming compressed sensing," Proc. IEEE International Symposium on Information Theory (ISIT), Cambridge, USA, pp. 1862-1866, 2012.
	
\bibitem{jian2011investigation}
B. Sun and J. Jiang, ``Investigation of sign spectrum sensing method," Acta Phys Sin-ch Ed, vol. 11, no. 32, pp. 110701, 2011.
	
\bibitem{biao2013fast}
B. Sun, Q. Chen, X. Xu, L. Zhang, and J. Jiang, ``A fast and accurate two-stage algorithm for 1-bit compressive sensing," IEICE Trans. Inf. \& Syst., vol. 96, no. 1, pp. 120-123, 2013.
	
\bibitem{gopi2013one}
S. Gopi, P. Netrapalli, P. Jain, and A. Nori, ``One-bit compressed sensing: Provable support and vector recovery," Proc. 30th International Conf. on Machine Learning (ICML), Atlanta, USA, pp. 154-162, 2013.
	
\bibitem{jacques2013robust}
L. Jacques, J. N. Laska, P. T. Boufounos, and R. G. Baraniuk, ``Robust 1-bit compressive sensing via binary stable embeddings of sparse vectors," IEEE T. Inform Theory, vol. 59, no. 4, pp. 2082-2102, 2013.
	
\bibitem{plan2013one}
Y. Plan and R. Vershynin, ``One-bit compressed sensing by linear programming," Commun Pur Appl Math, vol. 66, no. 8, pp. 1275-1297, 2013.
	
\bibitem{yan2012robust}
M. Yan, Y. Yang, and S. Osher, ``Robust 1-bit compressive sensing using adaptive outlier pursuit," IEEE T. Signal Proces,  vol. 60, no. 7, pp. 3868-3875, 2012.
	
\bibitem{plan2013robust}
Y. Plan and R. Vershynin, ``Robust 1-bit compressed sensing and sparse logistic regression: A convex programming approach," IEEE T. Inform Theory, vol. 59, no. 1, pp. 482-494, 2013.
	
\bibitem{ai2014one}
A. Ai, A. Lapanowski, Y. Plan, and R. Vershynin, ``One-bit compressed sensing with non-gaussian measurements," Linear Algebra Appl, vol. 441, pp. 222-239, 2014.
	
\bibitem{zhang2014efficient}
L. Zhang, J. Yi, and R. Jin, ``Efficient algorithms for robust one-bit compressive sensing," Proc. 31st International Conf. on Machine Learning (ICML), Beijing, China,  pp. 820-828, 2014.
	
\bibitem{trefethen1997numerical}
L. N. Trefethen and D. Bau, Numerical linear algebra, SIAM, Philadelphia, 1997.

\bibitem{li2015one}
P. Li, ``One scan 1-bit compressed sensing," arXiv preprint, arXiv: 1503.02346, 2015.

\bibitem{ma2014two}
Y. Ma, D. Baron, D. Needell, ``Two-Part Reconstruction with Noisy-Sudocodes," IEEE T. Signal Proces,  vol. 62, no. 23, pp. 6323-6334, 2014.
	
\end{thebibliography}

\appendix
\section{Proof of Theorem \ref{Theorem:APP}}
It is worth noting that
\begin{equation}
\begin{split}
& \mathbb{P}({\rm sign}(x^{\rm T}\phi) = {\rm sign}(\phi_j))\\
& = \mathbb{P}(x^{\rm T}\phi>0,\phi_j>0) + \mathbb{P}(x^{\rm T}\phi\leq0,\phi_j\leq0).
\end{split}
\end{equation}
We can divide $x^{\rm T}\phi$ into two parts as
\begin{equation}
m = x^{\rm T}\phi = m_j + m_c,
\end{equation}
where
\begin{equation}
\begin{split}
	m_j &= \phi_jx_j,\\
	m_c &= x^{\rm T}\phi-\phi_jx_j.\\
\end{split}
\end{equation}
In addition, it can be easily verified both $m_j$ and $m_c$ satisfy Gaussian distribution, i.e.,
\begin{equation}
\begin{split}
m_j &\thicksim \mathcal{N}(0,x_j^{2}),\\
m_c &\thicksim \mathcal{N}(0,1-x_j^{2}).\\
\end{split}
\end{equation}
Depending on $x_j$, we have three situations as follows,
(1) when $x_j=0$, we have
		\begin{equation}
		\begin{split}
		&\mathbb{P}(x^{\rm T}\phi>0,\phi_j>0)\\
		&=\mathbb{P}(m_c>0,m_j>0)\\
		&=\mathbb{P}(m_c>0)P(m_j>0)\\
		&=\frac{1}{4}.\\
		\end{split}
		\end{equation}
In the same way, we have
		\begin{equation}
			\mathbb{P}(x^{\rm T}\phi\leqslant0,\phi_j\leqslant0)=\frac{1}{4}.
		\end{equation}
Therefore,
		\begin{equation}
			\mathbb{P}({\rm sign}(x^{\rm T}\phi)={\rm sign}(\phi_j))= \frac{1}{2}.
		\end{equation}
(2) when $x_j>0$, we have
		\begin{equation}
			\mathbb{P}(x^{\rm T}\phi>0,\phi_j>0) = \mathbb{P}(m_c+m_j>0,m_j>0)
		\end{equation}
The joint probability density function of $m_c$ and $m_j$ is\\
		\begin{equation}
			p(m_c,m_j)=\frac{1}{2\pi x_j \sqrt{1-x_j^{2}}} \exp \left(-\frac{1}{2}\Big(\frac{m_j^{2}}{x_j^{2}}+\frac{m_c^{2}}{1-x_j^{2}}\Big)\right)
		\end{equation}
Assume that
		\begin{equation}
			\begin{split}
				m_c &= r\cos\theta,  \\
				m_j & =r\sin\theta, \\
			\end{split}
		\end{equation}
then we have,
\begin{equation}
\begin{split}
& \mathbb{P}(m_c+m_j>0,m_j>0) \\
&= \frac{1}{2\pi x_j\sqrt{1-x_j^{2}}}\int^{\frac{3}{4}\pi}_{0}\,d\theta \\
&\int^{\infty}_{0}\exp\left(-\frac{1}{2}\Big(\frac{r^{2}{\cos^{2}\theta}}{1-x_j^{2}}+\frac{r^{2}{\sin^{2}\theta}}{x_j^{2}}\Big)\right)rdr \\
&= \frac{1}{2}-\frac{1}{2\pi}\arccos(x_j). \\
\end{split}
\end{equation}
In the same way, we have
\begin{equation}
\begin{split}
&\mathbb{P}(x^{\rm T}\phi\leq0,\phi_j\leq0) \\
&= P(m_c+m_j\leq0,m_j\leq0) \\
&= \frac{1}{2}-\frac{1}{2\pi}\arccos (x_j). \\
\end{split}
\end{equation}
Therefore, we have
\begin{equation}
\mathbb{P}({\rm sign}(x^{\rm T}\phi)={\rm sign}(\phi_j))= 1-\frac{1}{\pi}{\rm arccos}(x_j).
\end{equation}
(3) when $x_j<0$,
\begin{equation}
\begin{split}
&\mathbb{P}({\rm sign}(x^{\rm T}\phi)={\rm sign}(\phi_j)) \\
& =\mathbb{P}(x^{\rm T}\phi>0,\phi_j>0)+\mathbb{P}(x^{\rm T}\phi\leq0,\phi_j\leq0) \\
& =\mathbb{P}(m_c+m_j>0,m_j<0)+\mathbb{P}(m_c+m_j\leq0,m_j\geq0).
\end{split}
\end{equation}
The first part can be computed via
\begin{equation}
\begin{split}
& \mathbb{P}(m_c+m_j>0,m_j<0) \\
&= \frac{1}{2\pi x_j\sqrt{1-x_j^{2}}}\int^{0}_{-\frac{1}{4}\pi}\,d\theta \\
& \int^{\infty}_{0}\exp\left(-\frac{1}{2}\Big(\frac{r^{2}}{1-x_j^{2}}{\cos^{2}\theta}+\frac{r^{2}}{x_j^{2}}{\sin^{2}\theta}\Big)\right)rdr \\
&= \frac{1}{2}-\frac{1}{2\pi}\arccos(x_j).\\
\end{split}
\end{equation}
In the same way, we calculate the second part as
\begin{equation}
\mathbb{P}(m_c+m_j<0,m_j>0)=\frac{1}{2}-\frac{1}{2\pi}\arccos (x_j).
\end{equation}
Therefore, we have
\begin{equation}
\mathbb{P}({\rm sign}(x^{\rm T}\phi)={\rm sign}(\phi_j))= 1-\frac{1}{\pi}{\rm arccos}(x_j).
\end{equation}
Synthesize the above three situations, we have
\begin{align}
\mathbb{P}({\rm sign}(x^{\rm T}\phi)={\rm sign}(\phi_j)) & = 1-\frac{1}{\pi}{\rm arccos}(x_j) \\
\mathbb{P}({\rm sign}(x^{\rm T}\phi)\neq{\rm sign}(\phi_j)) & =\frac{1}{\pi}{\rm arccos}(x_j).
\end{align}
This concludes the proof.

{\section{Proof of Lemma \ref{Lemma:sign-flip}}
In the noiseless setting, we define an event $E1$ to be
\begin{equation}
E1: \mathrm{sign}\left(y_i\right)\neq\mathrm{sign}\left(\mathbf{A}^i_j\right).
\end{equation}
From Theorem \ref{Theorem:APP}, we have
\begin{equation}
\mathbb{P}(E1) = \frac{1}{\pi}{\rm arccos}(x_j).
\end{equation}
In the noisy setting, we define an event $E2$ that $y_i$ has its sign flipped, i.e.,
\begin{equation}
E2: y_i = -1\cdot{\rm sign}(\mathbf{A}^i x),
\end{equation}
and by the definition of sign flip ratio, we have
\begin{equation}
\mathbb{P}(E2) = \rho.
\end{equation}
It is easy to verify that $E1$ and $E2$ are independent events, and we have
\begin{equation}
\begin{split}
P_j &= \mathbb{P}\left(\overline{E1}E2\right)+\mathbb{P}\left(E1\overline{E2}\right) \\
&= \mathbb{P}\left(\overline{E1}\right)\mathbb{P}\left(E2\right)+\mathbb{P}\left(E1\right)\mathbb{P}\left(\overline{E2}\right) \\
&= \left(1-\frac{1}{\pi}{\rm arccos}(x_j)\right)\rho+\frac{1}{\pi}{\rm arccos}(x_j)(1-\rho) \\
&= \frac{1-2\rho}{\pi}{\rm arccos}(x_j)+\rho.
\end{split}
\end{equation}
Then the proof completes.

\section{Proof of Lemma \ref{Lemma:error_bound}}
For any $x_u-x_v>\epsilon$, because $P_j$ is continuous on the closed interval $[x_v,x_u]$ and differentiable on the open interval $(x_v,x_u)$, by using the Lagrange's mean value theorem, there exists a point $x_c$ in $(x_v,x_u)$ such that
\begin{equation}
\label{apdx_lemma2:eq1}
\begin{split}
P_v-P_u &= \left.\frac{\partial P_j}{\partial x_j}\right|_{x_j=x_c}(x_v-x_u) \\
&= -\frac{(1-2\rho)}{\pi\sqrt{1-x_c^2}}(x_v-x_u) \\
&> \frac{(1-2\rho)\epsilon}{\pi}. \\
\end{split}
\end{equation}
Because both $H\{y,\mathbf{A}_u\}$ and $H\{y,\mathbf{A}_v\}$ obey the binomial distribution, i.e.,
\begin{equation}
\begin{split}
H\{y,\mathbf{A}_u\} \sim B(M,P_u), \\
H\{y,\mathbf{A}_v\} \sim B(M,P_v), \\
\end{split}
\end{equation}
therefore, the expectation and variance of $H\{y,\mathbf{A}_u\}$ and $H\{y,\mathbf{A}_v\}$ are given by
\begin{equation}
\begin{split}
\mathbb{E}\big(H\{y,\mathbf{A}_u\}\big) &= MP_u, \\
\mathbb{E}\big(H\{y,\mathbf{A}_v\}\big) &= MP_v, \\
\mathbb{D}\big(H\{y,\mathbf{A}_u\}\big) &= MP_u(1-P_u), \\
\mathbb{D}\big(H\{y,\mathbf{A}_v\}\big) &= MP_v(1-P_v). \\
\end{split}
\end{equation}
Define a random variable $z$ as
\begin{equation}
z \overset{\mathrm{def}}{=} H\{y,\mathbf{A}_v\}-H\{y,\mathbf{A}_u\}.
\end{equation}
Because $H\{y,\mathbf{A}_u\}$ and $H\{y,\mathbf{A}_v\}$ are independent, the expectation and the variance of $z$ are given by
\begin{equation}
\label{apdx_lemma2:eq2}
\begin{split}
\mathbb{E}\big(z\big) &= \mathbb{E}\big(H\{y,\mathbf{A}_v\}\big)-\mathbb{E}\big(H\{y,\mathbf{A}_u\}\big) \\
&= M(P_v-P_u), \\
\mathbb{D}\big(z\big) &= \mathbb{D}\big(H\{y,\mathbf{A}_v\}\big)+\mathbb{D}\big(H\{y,\mathbf{A}_u\}\big) \\
&= M\big(P_v(1-P_v)+P_u(1-P_u)\big).
\end{split}
\end{equation}
In addition, note that the probability mass function of $z$ is symmetrical with $\mathbb{E}\big(z\big)$. By using the Chebyshev's inequality, we have
\begin{equation}
\label{apdx_lemma2:eq3}
\begin{split}
\mathbb{P}(z>0) &= 1-\mathbb{P}(z\leq 0) \\
& = 1-\frac{1}{2}\mathbb{P}\big(\big|z-\mathbb{E}(z)\big|\geq\mathbb{E}(z)\big) \\
&\geq 1-\frac{\mathbb{D}(z)}{2\mathbb{E}^2(z)}. \\
\end{split}
\end{equation}
Let $f(z)=1-\frac{\mathbb{D}(z)}{2\mathbb{E}^2(z)}$, by substituting (\ref{apdx_lemma2:eq1}) and (\ref{apdx_lemma2:eq2}) into $f(z)$, we have
\begin{equation}
\begin{split}
&f(z)= \\
&1+\frac{1}{2M}+\frac{\pi^2P_u(P_u-1)}{M\epsilon^2(1-2\rho)^2}+\frac{\pi(P_u(2-4\rho)+2\rho-1)}{2M\epsilon(1-2\rho)^2} \\
\end{split}
\end{equation}
By computing the derivative of $f(z)$ with respect to $P_u$ and set it to be $0$, we compute the minimum value of $f(z)$ to be
\begin{equation}
\label{apdx_lemma2:eq4}
f(z)\geq 1-\frac{\pi^2-\epsilon^2(1-2\rho)^2}{4M\epsilon^2(1-2\rho)^2}.
\end{equation}
By combining (\ref{apdx_lemma2:eq3}) and (\ref{apdx_lemma2:eq4}), we have
\begin{equation}
\mathbb{P}\big(H\{y,\mathbf{A}_v\}>H\{y,\mathbf{A}_u\}\big)\geq 1+\frac{1}{4M}-\frac{\pi^2}{4M(1-2\rho)^2}\epsilon^{-2}.
\end{equation}
Then the proof completes.}

\clearpage
\profile[BiaoSun]{Biao Sun}{received the Ph.D. degree in electrical science \& technology from Huazhong University of Science and Technology in 2013. He is currently an assistant professor with the Department of Electrical Engineering \& Automation, Tianjin University, China. His research interests include neural information processing, compressed sensing, and machine learning.
}
%
\profile[HuiFeng]{Hui Feng}{is a master's degree in the Department of Electrical Engineering \& Automation, Tianjin University, China. She is interested in compressed sensing.
}
\profile[XinxinXu]{Xinxin Xu}{
received the Ph.D. degree in electrical science \& technology from Huazhong University of Science and Technology in 2013. She is currently an engineer with the Microsystems Technology Center, Information Science Academy of China Electronics Technology Group Corporation. Her research interests include meta material, frequency selective surfaces,electromagnetic protection and electromagnetic measurement.
}

\end{document}